\documentclass{article}
\usepackage{spconf,amsmath,graphicx}

\usepackage{times}  
\usepackage{helvet}  
\usepackage{courier}  
\usepackage{url}  
\usepackage{graphicx}  
\usepackage{amsthm}
\usepackage{diagbox}
\usepackage{multirow}
\usepackage{enumerate}
\usepackage{epsfig}
\usepackage{algorithmic}
\usepackage{algorithm}
\usepackage{amsmath}
\newcommand{\keypoint}[1]{\vspace{0.1cm}\noindent\textbf{#1}\quad}
\newcommand{\tb}[1]{\textbf{#1}}

\title{Deep Stock Representation Learning: From Candlestick Charts to Investment Decisions}

\name{\parbox{0.6\linewidth}{\centering\footnotesize{Guosheng Hu$^{5,*}$\thanks{$^*$ These authors contributed equally to this work} Yuxin Hu$^{1,*}$ Kai Yang$^2$ Zehao Yu$^3$ Flood Sung$^4$ Zhihong Zhang$^3$ Fei Xie$^{1,\diamondsuit}$ \thanks{ $^\diamondsuit$ Corresponding author: xiefei@mail.shufe.edu.cn} Jianguo Liu$^1$ Neil Robertson$^{5}$ Timothy Hospedales$^{6,\heartsuit}$ Qiangwei Miemie$^{6,7,8,\heartsuit}$\thanks{$^\heartsuit$ Email: t.hospedales@ed.ac.uk, yongxin.yang@arraystream.com} }}\vspace{-0.5em}}
\address{\parbox{0.85\linewidth}{\centering\scriptsize{$^1$Shanghai University of Finance and Economics~ $^2$University of Shanghai for Science and Technology~ $^3$Xiamen University~ $^4$Independent Researcher $^5$Queen's University Belfast~ $^6$The University of Edinburgh~ $^7$Yang's Accounting Consultancy Ltd~ $^8$ArrayStream Technologies Ltd}}}

\usepackage[small,compact]{titlesec}

\begin{document}
\maketitle

\begin{abstract}

We propose a novel investment decision strategy (IDS) based on deep learning. 
The performance of many IDSs is affected by stock similarity. 
Most  existing stock similarity measurements have the problems: 
(a) The linear nature of many measurements  cannot capture nonlinear stock dynamics; 
(b) The estimation of many similarity metrics (e.g. covariance) needs very long period historic data (e.g. 3K days) which cannot represent current market effectively; 
(c) They cannot capture translation-invariance. 
To solve these problems,  we apply Convolutional AutoEncoder  to learn a stock representation, 
based on which we propose a novel portfolio construction strategy by: (i) using the deeply learned representation and modularity optimisation to cluster stocks and identify diverse sectors,
(ii) picking stocks within each cluster according to their Sharpe ratio (Sharpe 1994).
Overall this strategy provides low-risk high-return portfolios. We use the Financial Times Stock Exchange 100 Index (FTSE 100) data for evaluation.
Results show our portfolio outperforms FTSE 100 index and many well known funds in terms of total return in 2000 trading days.

\end{abstract}

\section{Introduction}

Investment decision making  is a classic research area in quantitative and behavioural finance.
One of the most important decision problems is portfolio construction and optimisation \cite{markowitz1952portfolio,kelly1956new}, 
which addresses selection and weighting of assets to be held in a portfolio. 
Financial institutions try to construct and optimise portfolios in order to maximise investor returns while minimising investor risk.

Stock similarity is important for many investment decision strategies ~\cite{ledoit2003improved}.  
For example, the classical investment strategy, mean-variance theory \cite{markowitz1952portfolio}, measures stock similarity using variance. Most  similarity measurements have the following problems: 
(a) Usually, the time series (linear  signal) is fed to linear  metric  (e.g. covariance, Pearson) to obtain similarity. 
The linear nature of most  similarities cannot capture the nonlinear dynamics of the stocks. 
(b) In \cite{chamberlain1982arbitrage}, it is claimed  that $n$ (the number of stocks in one market, e.g. 2,033 tradable stocks in London Stock Exchange) 
historic days/weeks data is needed to estimate an accurate covariance. However, the past $n$ days/weeks data cannot represent the current market effectively. 
(c) Most  similarity measurements do not consider translation (time)-invariance, which is important for stock similarity. For example, the price of  Apple stock increases at one particular day, however, 
the stock prices of Apple suppliers might increase after 3 days.

To solve the  aforementioned problems,   we propose to use deep learning (DL) features for stock similarity measurement instead of raw time series. 
Convolutional DL approaches such as  Convolutional AutoEncoder (CAE, unsupervised) \cite{masci2011stacked} and  Convolutional Neural Network (CNN, supervised) \cite{lecun1998gradient}, 
have achieved very impressive performance for analysing visual imagery. 
This has motivated researchers to convert raw input signals from other modalities into images to be processed by CNNs or CAEs. 
In this way, good results have been achieved for diverse applications. 
For example, traditional speech recognition methods  used the 1-D signal vector, e.g. the raw input waveform \cite{rabiner1993fundamentals,povey2011kaldi}. 
In contrast an alternative approach is to convert the 1-D signal to a  spectrogram, i.e. an  image, in order to leverage the strength of CNNs to achieve promising recognition performance \cite{amodei2016deep}. 
As another well known example, AlphaGo ~\cite{silver2016mastering}  represents the board position as a 19$\times$19 image, which is fed into a CNN for feature learning. Besides, computer vision techniques have also been applied to judge the quality of paper \cite{Bearnensquash2010} and calculate the rank of matrix \cite{Fouhey2015} from its appearance only.   
With similar motivation, we  explore to convert
a 4-channel stock time-series (lowest, highest, opening and closing price for the day) to candlestick charts by synthesis technique to present price history as images. 
To avoid expensive annotation, we choose the  unsupervised CAE for stock feature learning using the synthetic candlestick images.

Hence, the first novelty of this study is exploiting deep learning (i.e. CAE) to encode stock time series. 
Compared with raw time series, deeply learned features can effectively capture (i)   nonlinear stock dynamics and semantics; (ii)  the translation-invariance. 
The similarity measurements based on deep features can overcome the aforementioned weaknesses of most existing measurements. 
In addition, we contribute a new valuable signal, deep feature, to the  investment decision society, in which new effective signal is important for risk hedging. 
Though some deep learning models, such as LSTM ~\cite{fischer2017deep} and RNN ~\cite{singh2017stock}, have been applied  to optimise portfolio, 
they use raw time series rather than charts as input.

Second, motivated by momentum effect \cite{jegadeesh1993returns},  we  construct a novel portfolio generation pipeline including: (1) deep feature learning by visual interpretation price history, 
(2) clustering the stocks based on the similarity computed on deep features to provide a data-driven segmentation of the market, 
(3) actual portfolio construction. 
For \emph{visual representation} learning,  we generate millions of training images (synthetic candlestick charts) which are fed to a deep CAE for feature learning.  
In the next \emph{clustering} step, we aim to segment the market into diverse sectors in a data-driven way.  
This is important to provide risk reduction by selecting a well diversified portfolio \cite{nanda2010clustering,tola2008cluster}. 
The similarity embedded in clustering method is computed using deep features.  
Popular clustering methods such as K-means are not suitable here because they are non-deterministic and/or require a pre-defined numbers of clusters to find. 
In particular non-deterministic methods are not acceptable to real financial users. 
To address this we adapt the modularity optimization method \cite{newman2006modularity} -- originally designed for network community structure -- to stock clustering. 
Finally, we perform \emph{portfolio construction} by the simple yet effective approach of choosing the best stock within each cluster according to their  Sharpe ratio \cite{sharpe1994sharpe}. 
As we will see in our evaluation, this    portfolio selection strategy  combines high returns with low risk.

\section{Methodology}
\begin{figure}[t]
\begin{center}		
\includegraphics[trim = 15mm 60mm 50mm 20mm, clip, width=1 \linewidth]{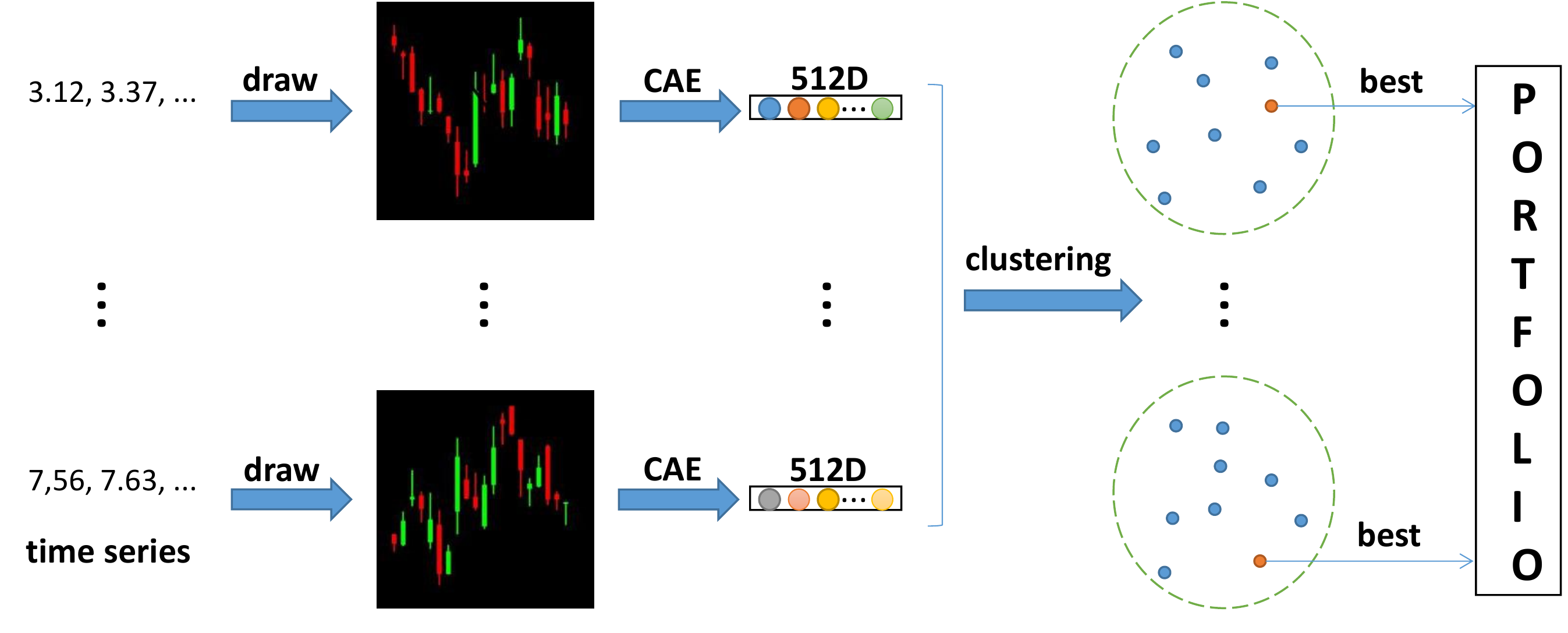}		
\end{center}
\vspace{-2em}\caption{\small Schematic illustration of our investment decision pipeline. The architecture of CAE is detailed in Fig.~\ref{fig:arch}. }	
\label{fig:pipeline}	
\end{figure}

Our overall investment decision pipeline includes three main modules: deep feature learning,  clustering, and portfolio construction.  
For \emph{deep feature learning},  raw 4-channel time series data describing stock price history are converted to standard candlestick charts. 
These charts  are  fed into deep CAEs for visual feature learning. These learned features provide a 
{vector embedding} of a historical time-series that captures key quantitative and semantic information. 
Next we \emph{cluster} the features in order to provide a data-driven segmentation of the market to underpin subsequent selection of a diverse portfolio. 
Many common clustering methods are not suitable here because they are non-deterministic or require predefinition of the number of clusters. 
Thus we adapt modularity optimisation for this purpose. 
Note that stock similarity embedded in our clustering method is computed using the nonlinear deep features. 
Finally, we perform \emph{portfolio construction} by choosing stocks with the best performance measured by Sharpe ratio \cite{sharpe1994sharpe} from each cluster. The overall pipeline is summarised schematically in Fig.~\ref{fig:pipeline}. Each component is discussed in more detail in the following sections.

\subsection{Deep Feature Learning with CAEs}
\label{sec:dfl}

\begin{figure}[t]
\vspace{-1em}
\begin{center}		
\includegraphics[trim = 15mm 50mm 36mm 35mm, clip, width=0.95 \linewidth]{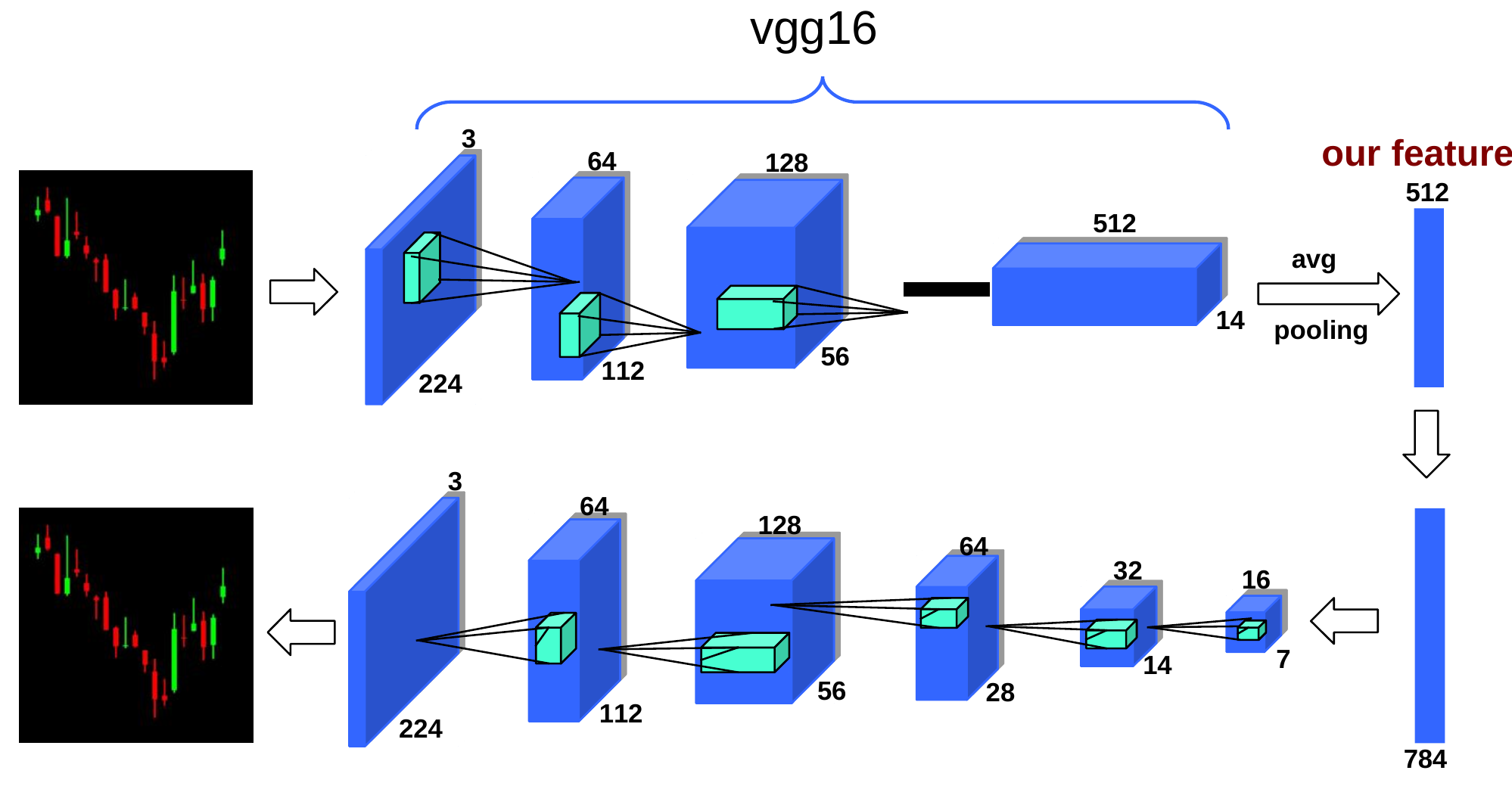}		
\end{center}
\vspace{-2em}\caption{\small CAE overview. An encoder (top) - decoder (bottom) framework. 
The 512D feature following average pooling provides our representation for clustering and portfolio construction.  }	
\label{fig:arch}
\vspace{0.2em}	
\end{figure}

\keypoint{Chart Encoding}
To realise an algorithmic portfolio construction method based on visual interpretation of stock charts, we need to convert raw price history data to an image representation. 
Our raw data for each stock is a 4-channel time series (the lowest, the highest, open, and closing price for the day) in a 20-day time sequence. We use computer graphics techniques to convert these to a candlestick chart represented as 
a RGB image as shown in Fig.~\ref{fig:pipeline} and ~\ref{fig:arch}. The whisker plots describe the four raw channels, with colour coding describing whether the stock closed higher (green) or lower (red) than opening. 
An encoded candlestick chart image provides the visual representation of one stock over a 20-day window for subsequent visual interpretation by our deep learning method. 

\keypoint{Convolutional Autoencoder}
Our CAE architecture is summarised in  Fig.~\ref{fig:arch}. It is based on the landmark VGG network \cite{simonyan2014very}, specifically VGG16. The VGG network is a highly successful architecture initially proposed for visual recognition. To adapt it for use as a CAE encoder, we remove the final  4096D FC layers from VGG-16 and  replace them by an average pooling layer to generate one 512D feature.
The decoder is a 7-layer deconvolutional network that starts with a 784D layer that is fully connected with the 512D embedding layer. Following  6 up-sampling deconvolution layers eventually reconstruct the input based on our 512D feature. 
When trained with a reconstruction objective, the CAE network learns to compress input images to the 512D bottleneck in a manner that preserves as much information as possible in order to be able to reconstruct the input. Thus  this single 512D vector encodes the 20-day 4-channel price history of the stock, and will provide the representation for further processing (clustering and portfolio construction). 

\subsection{Clustering}

We next aim to provide a clustering method for diversified -- and hence low risk -- portfolio selection. 
As discussed, many existing clustering methods are non-deterministic or require pre-specification of the number of clusters, which make them unsuited for our application.
To solve these problems, we introduce the network modularity method \cite{newman2006modularity} to find the cluster structure of the stocks, 
where each stock is set as one node and the link between each pair of stocks is set as the cosine similarity calculated by our learned  CAE features. 
Modularity is introduced as the  fraction of the links that fall within the given group minus the expected fraction if links are distributed at random. 
Modularity optimisation \cite{newman2006modularity}, originally used for detecting community structure in networks, can end with generating clusters. 
Specifically, optimisation operates on a graph (one 20-day history of the entire market in our case), and updates the graph to group stocks so as to eventually achieve maximum modularity before terminating. 
Thus it does not need a specified number of clusters and is not affected by initial node selection.

\subsection{Portfolio Construction and Backtesting}
\label{sec:pcab}
Given the learned stock clustering (market segmentation) , we construct a complete portfolio  by picking diverse yet high-return stocks, and evaluate the result.

\keypoint{Stock Performance} 
Return (profit on an investment) is defined as $r_t=(V_{f}-V_{i})/V_{i}$, where $V_{f}$ and $V_{i}$ are the final and initial values, respectively. 
For example, to compute daily stock return,  $V_{f}$ and $V_{i}$ are closing prices of today and yesterday, respectively.
We measure the performance of one particular stock over a period using the Sharpe ratio \cite{sharpe1994sharpe} $s=\overline{r}/\sigma_r$, where $\overline{r}$ is the mean return, 
$\sigma_r$ is the standard deviation over that period. Thus the Sharpe ratio $s$ encodes a trade-off of return and stability. 
Maximum Drawdown (MDD) is the measure of decline from peak during a specific period of investment: $\text{MDD}=(V_{t}-V_{p})/V_{p}$, where $V_{t}$ and $V_{p}$ mean the trough and peak values, respectively.

\keypoint{Training and Testing}
For every 20 trading days, we cluster all the stocks. 
To actually construct a portfolio we then choose the stock with the highest Sharpe ratio \cite{sharpe1994sharpe} within each cluster. 
We then hold the selected portfolio for 10 days. Over these following 10 days, we evaluate the portfolio by computing our `compound return'  for each selected stock. 
The overall return of one portfolio  is the average compound return of all the selected stocks. We use a stride of 10.
The process of portfolio selection and return computation are analogous to  training and testing process in machine learning, respectively.  

\keypoint{Fund Allocation}
Since our clustering method discovers the number of stocks in a data driven way,  in different trading periods we may have different number of clusters. 
Assume that we obtain $K_1$ clusters in one period and will select $K_2$ stocks to construct one portfolio. 
Then, letting $Q$ and $R$ indicate quotient and remainder respectively in  $[Q, R]=K_2/K_1$: $K_2$ stocks are picked by taking  (i) $Q$ stocks from each  of the $K_1$ clusters 
and (ii) the remaining $R$ best performing stocks {across all} $K_1$ clusters. Then, we allocate \emph{equally} $1/K_2$ of the fund to each of the chosen  stocks.

\section{Experiments}
We first introduce our dataset and experiment settings. We analyse the outputs of feature learning. 
Finally, we compare our whole investment strategy (feature extraction, clustering, portfolio optimisation) with alternatives.

\subsection{Dataset and Settings}
For evaluation, we  use the stock data of Financial Times Stock Exchange 100 Index (FTSE 100), which is a share index of the 100 companies listed on the London Stock Exchange with the highest market capitalisation. 
We use all the stocks in  FTSE 100   from  4th Jan 2000 to 14th May 2017. 
The stock price is adjusted accounting for stock splits, dividends and distributions.
Every 20-day 4-channel  time series generates a standard candlestick chart.  
We generate 400K FTSE100  charts in all. The training images for our CAE are candlestick charts rendered as 224 $\times$ 224 images to suit our VGG16 architecture \cite{simonyan2014very}. 
During training, the batch size is 64, learning rate is set to 0.001, and the learning rate decreases with a factor of 0.1 once the network converges.

\subsection{Qualitative Results}
\begin{figure}[t]
\small
\begin{center}		
\includegraphics[trim = 45mm 15mm 40mm 20mm, clip, width=0.95 \linewidth]{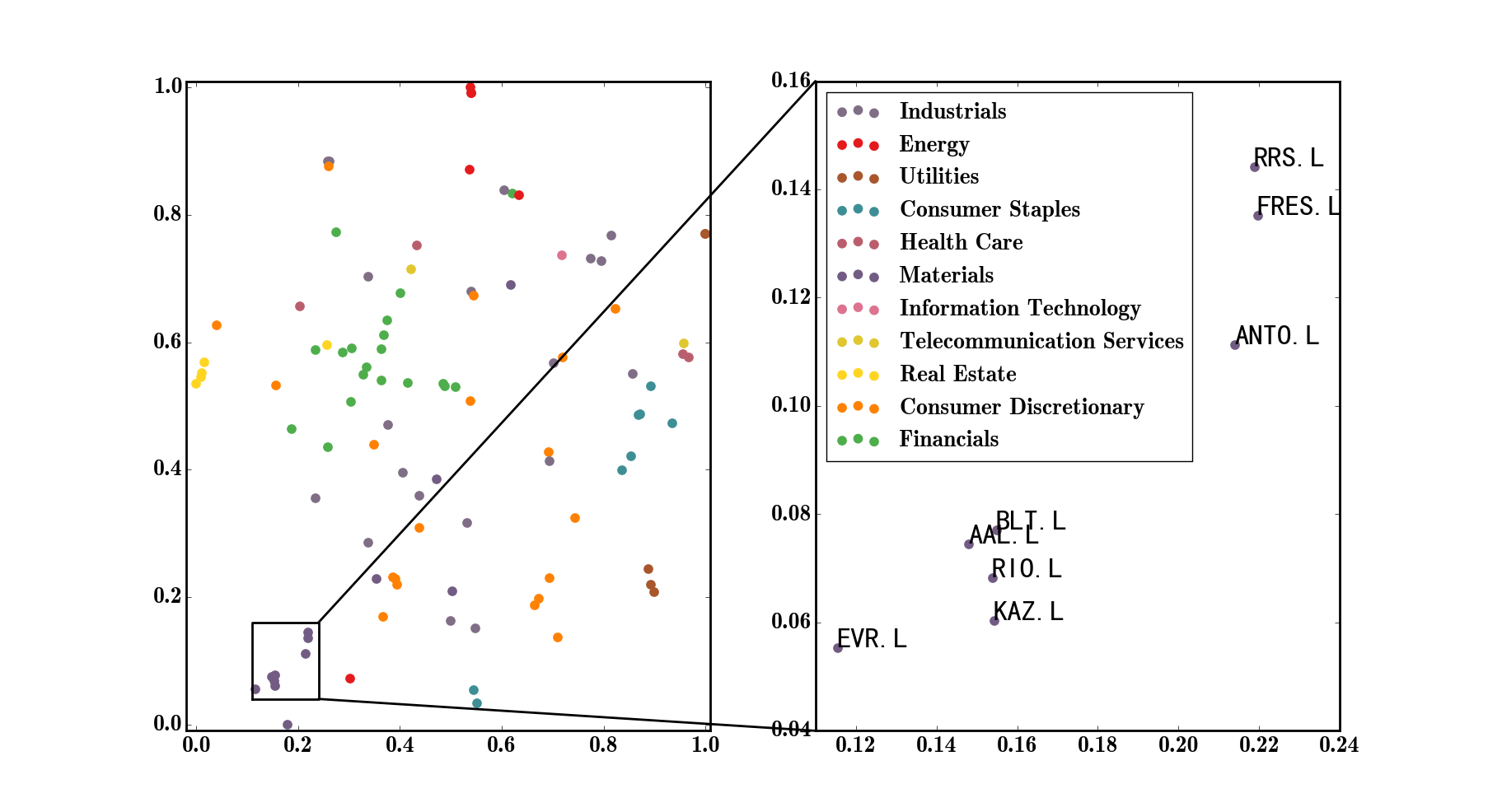}		
\end{center}
\vspace{-2em}\caption{\small {t-SNE visualisation of FTSE 100 CAE features. One colour indicates one industrial sector. 
The stocks on the right are all from Sector Materials: RRS.L (Randgold Resources Limited), FRES.L (Fresnillo PLC), ANTO.L (Antofagasta PLC), BLT.L (BHP Billiton Ltd), AAL.L (Anglo American PLC), RIO.L (Rio Tinto Group), KAZ.L (KAZ Minerals), EVR.L (EVRAZ PLC)}}	
\label{fig:fv}
\end{figure}

\keypoint{Visualising and Understanding Deep Features} 
The features of one year (2012) for a given stock are concatenated to form a new feature. 
These features of all the stocks are visualised in Fig.~\ref{fig:fv} using the t-Distributed Stochastic Neighbour Embedding (t-SNE) \cite{maaten2008visualizing} method. 
One colour indicates one industrial sector defined by Bloomberg
From Fig.~\ref{fig:fv}, we can see the stocks with similar semantics (industrial sector) are represented close to each other in the learned feature space. 
For example, Materials related stocks are clustered. This illustrates the efficacy of our CAE and learned feature for capturing semantic information about stocks.

\subsection{Quantitative Results}
For quantitative evaluations, we apply 7 measures for evaluation: Total return, daily Sharpe ratio, max drawdown, daily/monthly/yearly mean return, and win year. 
Win year indicates the percent of the winning years. The other measures are defined in the section of `Portfolio Construction and Backtesting'. 
{We choose $K_2=5$ stocks to construct all the portfolios compared.}

\begin{figure}[t!h]
\begin{center}		
\includegraphics[ trim = 100mm 20mm 90mm 20mm, clip, width=0.99 \linewidth]{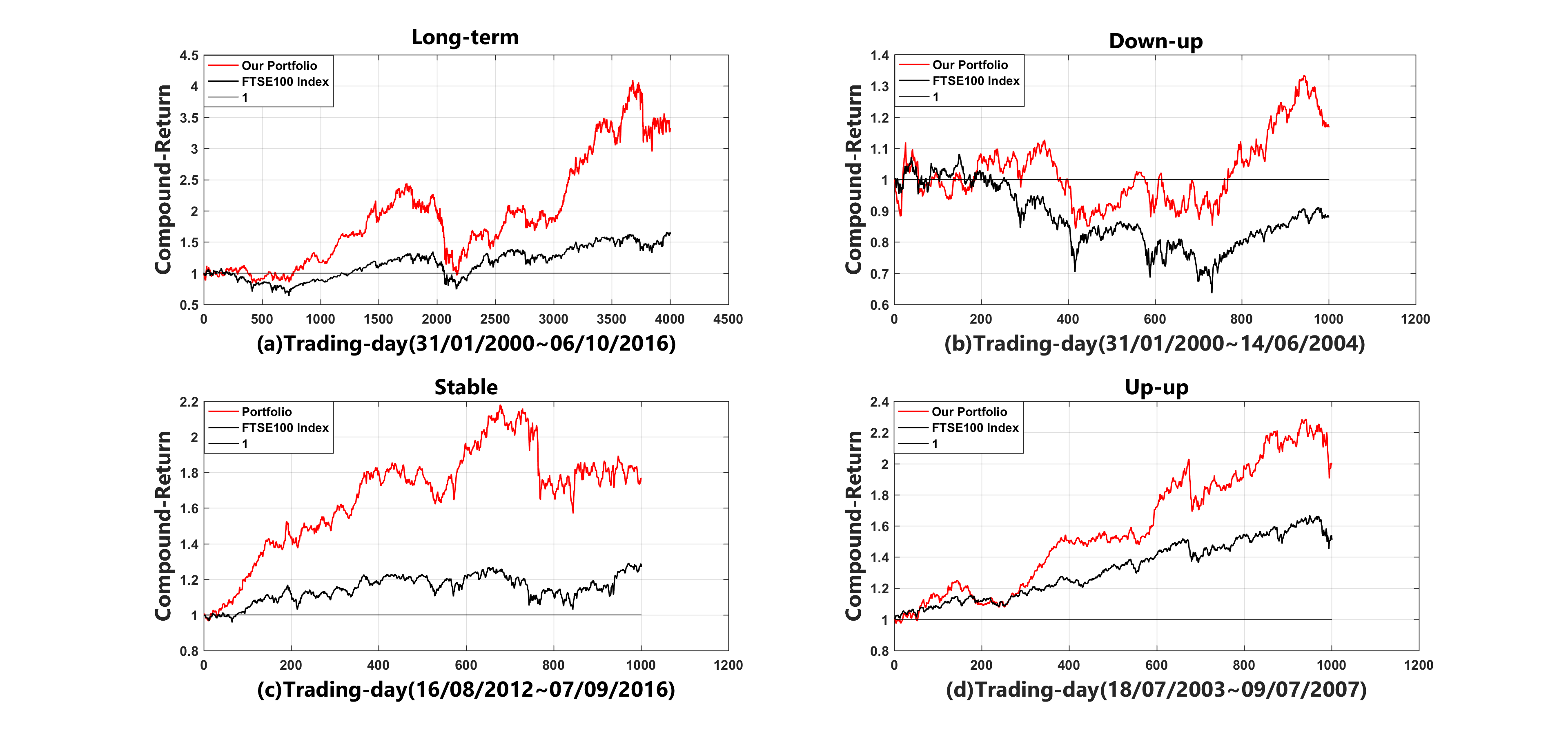}		
\end{center}
\vspace{-2em}\caption{\small Our portfolio vs FTSE 100 Index}	
\label{fig:fm}
\vspace{0em}
\end{figure}

\keypoint {Comparison with FTSE 100 Index} 
We perform backtesting to compare our full portfolio optimisation strategy against the market benchmark (FTSE 100 Index). 
In Fig. ~\ref{fig:fm}, we compare with FTSE 100 index.  Fig.~\ref{fig:fm} (a) shows the comparison over a long-term trading period (4K trading days, 31/01/2000 - 06/10/2016) showing the overall effectiveness of our strategy. 
Note that it is very difficult for funds to consistently outperform the market index over an extended period of time due to the complexity and diversity of market variations. 
In Fig.~\ref{fig:fm} (b)-(c) we show specific shorter term periods where the market is behaving very differently 
including down-up (b), flat (c) and bullish (d). 
The overall dynamic trends of our strategy reflect the conditions of the market (meaning that the stocks selected by our strategy are representative of the market), 
yet we outperform the market even across a diverse range of conditions (b-c), and over a long time-period (a).

\keypoint{Feature and Clustering} We evaluate features and clustering methods over a  long term period (4K trading days). 
From Table~\ref{tab:fac}, the total return of our method (D-M, deep feature + modularity-based clustering) is higher than R-M (R-M, Raw time series + modularity-based clustering),  
283.5\% vs 208.8\%. It means the deeply learned feature capture richer information, which is more effective for portfolio optimisation than raw time series. 
Similar conclusions can be drawn based on other measures. 
In terms of clustering method, our modularity optimization method works better than D-K (deep feature + $k$-means) in terms of returns and daily Sharpe, showing the effectiveness of modularity-based clustering. 
As explained in the Introduction, $k$-means cannot be used for portfolio construction in practice. 
Specifically, the results of  $k$-means cannot be repeated because of the  randomness of the initial seed. 
Non-deterministic investment strategies are not acceptable to financial users in practice as they add another source of uncertainty (risk) that is hard to quantify.

\begin{table}[t]
\centering
\caption{Comparison of Features and Clustering Methods.}
\label{tab:fac}
\small
\resizebox{0.45\textwidth}{!}{%
\begin{tabular}{cccc}
\hline
& R-M     & \tb{D-M (Ours)}         & D-K    \\ \hline 
Total Ret.  ($\uparrow$) & 208.8\%  & \tb{283.5\%} & 272.6\% \\ 
Daily Sharpe ($\uparrow$)   & 0.44  &  \tb{0.50} &  0.49      \\ 
Max Drawdown ($\uparrow$) & \tb{-55.6\%}  & -60.5\% & -59.0 \%\\ 
Daily Mean Ret. ($\uparrow$)   & 9.7 \%  & \tb{11.1\%} &  10.9\%  \\ 
Monthly Mean Ret. ($\uparrow$)  & 8.7\%  &\tb{10.0} \% & 9.9\%  \\ 
Yearly Mean Ret.  ($\uparrow$)  & 9.6\%  & 10.0 \% & \tb{11.19\%}   \\ 
Win Years  ($\uparrow$)   & 64.71\%  & \tb{69.52\%} & 66.31\%  \\ \hline
\end{tabular}%
}
\end{table}

\keypoint{Comparison with Funds}
To further analyse the effectiveness of our strategy, we compare our strategy with well known public funds in stock market in Table ~\ref{tab:cwf}. 
Specifically, we select 2  big funds (CCA and VXX) and {the top} 3 best performed funds (IEO, PXE, PXI)  recommended by YAHOO ( \url{https://finance.yahoo.com/etfs}). 
Note that the ranking of funds change over time. The fund data is obtained from Yahoo Finance.  Because VXX starts from 20/01/2009, this evaluation is computed over 2K trading days (20/01/2009-09/01/2017). 
From Table ~\ref{tab:cwf}, our portfolio achieved the highest returns: Total (215.4\%), daily (16.7\%), monthly (16.6\%), yearly (11.8\%) in 2000 trading days, 
showing the strong profitability of our strategy. We also achieved the  highest daily Sharpe ratio (0.8), meaning that we effectively balance the profitability and variance. 
We achieve the 2nd lowest max drawdown, meaning that our method can effectively manage the investment risk.  In most years (62.5\%), our portfolio makes a profit. 
It is only slightly worse than PXI in terms of 75.0\% of profitable years. This shows the stability of our strategy.

\begin{table}[h]
\vspace{-1.5em}
\centering
\caption{Comparison with Well-known Funds}
\label{tab:cwf}
\resizebox{0.45\textwidth}{!}{%
\begin{tabular}{ccccccc}
\hline
& CCA     & VXX     & IEO     & PXE     & PXI     & \textbf{Ours}    \\ \hline 
Total Ret.    & 117.0\% & -99.9\% & 89.9\%  & 101.6\% & 152.2\% & \textbf{215.4\%} \\ 
Daily Sharpe   & 0.7     & -1.1    & 0.4     & 0.4     & 0.6     & \textbf{0.8}     \\ 
Max Drawdown  & \textbf{-22.2\%}  & -99.9\% & -56.8\% & -57.6\% & -59.3\% & -30.9\% \\ 
Daily Mean Ret.    & 10.9\%  & -67.7\% & 12.7\%  & 13.4\%  & 15.9\%  & \textbf{16.7\%}  \\ 
Monthly Mean Ret.    & 10.7\%  & -66.4\% & 11.5\%  & 12.4\%  & 14.9\%  & \textbf{16.6\%}  \\ 
Yearly Mean Ret   & 6.5\%   & -44.6\% & 5.0\%   & 8.2\%   & 9.4\%   & \textbf{11.8\%}  \\ 
Win Years     & 62.5\%  & 0.0\%   & 62.5\%  & 50.0\%  & \textbf{75.0}\%  & 62.5\%  \\ \hline
\end{tabular}%
}
\vspace{-0.5em}
\end{table}

\section{Conclusions}
We propose a deep learned-based investment strategy, which includes: (1) novel stock representation learning by deep
CAE encoding of candlestick charts, (2) diversification through modularity optimisation based clustering and (3)
portfolio construction by selecting the best Sharpe ratio stock in each cluster. Experimental results show: (a) our
learned stock feature captures semantic information and (b) our portfolio outperforms the FTSE 100 index and many
well-known funds in terms of total return. 

\keypoint{Acknowledgements}
This work was supported by EPSRC (EP/R026173/1), the European Union's Horizon 2020 research and innovation program under grant agreement No 640891, and National Natural Science Foundation of China No. 61773248.
\newpage

\end{document}